# Application of Text Analytics in Public Service Co-Creation: Literature Review and Research Framework

Application of Text Analytics in Public Service Co-Creation

Nina Rizun

Gdańsk University of Technology, Fahrenheit Universities, Poland, nina.rizun@pg.edu.pl

Aleksandra Revina

Brandenburg University of Applied Sciences, Germany, revina@th-brandenburg.de

Noella Edelmann

University for Continuing Education Krems, Austria,  noella.edelmann@donau-uni.ac.at

The public sector faces several challenges, such as a number of external and internal demands for change, citizens' dissatisfaction and frustration with public sector organizations, that need to be addressed. An alternative to the traditional top-down development of public services is co-creation of public services. Co-creation promotes collaboration between stakeholders with the aim to create better public services and achieve public values. At the same time, data analytics has been fuelled by the availability of immense amounts of textual data. Whilst both co-creation and TA have been used in the private sector, we study existing works on the application of Text Analytics (TA) techniques on text data to support public service co-creation. We systematically review 75 of the 979 papers that focus directly or indirectly on the application of TA in the context of public service development. In our review, we analyze the TA techniques, the public service they support, public value outcomes, and the co-creation phase they are used in. Our findings indicate that the TA implementation for co-creation is still in its early stages and thus still limited. Our research framework promotes the concept and stimulates the strengthening of the role of Text Analytics techniques to support public sector organisations and their use of co-creation process. From policy-makers' and public administration managers' standpoints, our findings and the proposed research framework can be used as a guideline in developing a strategy for the designing co-created and user-centred public services.

## 1 INTRODUCTION

Given dramatic technology developments, the need to fulfil public values such as effective and efficient use of resources, citizens' lack of trust in governments, and satisfaction with public services [88, 96], public administrations are trying to find new ways of developing and delivery public services. One such way is the participation of citizens in the development and production of public services, i.e., *co-creation*. Although seen as a form of modernization in current public administration and management, co-creation is not a new concept in governance. The concept was developed in the 1970s by Ostrom to encourage and explain the role of citizens in the production of public services [78]. Co-creation reflects a move away from the New Public Management (NPM), a paradigm based on founded on disaggregation, competition, and incentivization for reforming government institutions [32] and a move towards a Public Service Logic (PSL) perspective [77]. Osborne points out that "*PSL [...] starts from the service user as its basic unit of analysis and explores how public services and PSOs [Public Sector Organisations] might be designed to facilitate the co-creation of value by service users, not vice versa [...] the potential to dramatically reconstruct how we conceptualize and govern the creation of value through the delivery of public services*" [77] (p. 229). With PSL, public administrations recognize and integrate citizens as essential stakeholders in the development and delivery of public services [18].

Whilst scholars such as Osborne note that citizens are always involved in the delivery of services [77], other scholars have also been interested in citizens' involvement in the development of the services [17], particularly in the context of the digital transformation of public administrations and the opportunities for involvement using digital tools and channels, e.g., [56]. Although IT and digital tools were not seen as particularly important for communication and involvement with citizens and other stakeholders, digital platforms and social media are now central in the management of public administrations as they allow for synchronous and interactive communication, accessibility and participation [11, 24]. At the same time, the use of these tools leads to the generation of huge amounts of digital textual data every minute [95]. Whilst some argue that this leads to information overload and a lack of transparency [3], Text Analytics is a technology that can be used to deal with the copious amounts of text data to extract meaning from them in a systematic and useful way [8].

Text Analytics (TA) began in the 1940s with the development of Content Analysis, Computational Linguistics, and Natural Language Processing [5]. The challenge of extracting meaningful insights from data, knowledge from structured data and unstructured texts began to emerge in the 1980-1990s [35]. Since then, data and text mining have been gaining popularity. In the era of Big Data in the 2010s and the extensive use of corporate data by a wide range of businesses, applied TA solutions began to appear [5]. Whilst in the past TA was only used to a limited extent in the public service sector [68], the interest in TA applications such as chatbots [93], public opinion analyses [66], and expert systems [9] has increased. This makes Text Analytics one of the most promising technologies to support digital co-creation in public administrations, find new services (cases) that could benefit from co-creation, thus contribute to increasing service quality, citizen satisfaction with services offered and trust in public sector organizations. With this study, we aim to answer the following research questions:

RQ1: *During which co-creation phases can TA support the development of public services?*

RQ2: *Which TA techniques best support digital co-creation?*

RQ3: *Which public service values do TA techniques contribute to co-creation?*

In order to answer these questions, the current study is based on a structured literature analysis of TA research and its applications in public services.

In the following section, we present the theoretical background on co-creation and Text Analytics. This is followed by the methodology, including the processes for data collection and analysis, the review findings and theoretical research framework on TA application for public service co-creation. We conclude this study with the limitations we identified, as well as some suggestions for future research.

## 2 TEXT ANALYTICS AND PUBLIC SERVICE CO-CREATION

The development of TA methods and applications has increased given the growing amount of digital textual data, the promising capabilities of Information Systems, and computing power to capture, process, and store data [5]. TA solutions support textual data processing and interpretation while delivering faster and better-quality results [74]. Some of the most common techniques are machine learning applied to text, e.g., text summarization, classification, and clustering [5], sentiment analysis [58], topic modeling [15], semantic approaches such as ontologies and knowledge graphs [61], artificial neural networks like word embedding models [71], information extraction, multilingual solutions [45]. As an outcome, several practical use cases, such as email filtering, product and service ratings and recommendations, opinion mining, trend analysis, and search engines, have been established [5]. It will take years for technology acceptance and dissemination to impact organizational performance, processes, and legal and cultural aspects [21]. In the particular context of the public sector, there are additional constraints to the implementation of digital innovations, such as lack of understanding its importance, lack of competences, data protection requirements and disaggregation between the public sector organizations as a result of NPM [68]. Thus, despite the advancements in TA and the development of the numerous TA techniques, our understanding of how TA affects the labour force, the economy, governance, or society at large is fragmented and lacks a joint theoretical guidance [48, 68]. In addition, current research and knowledge about the TA techniques, functionalities, values, and potential of applications in the public sector need to be systematized. Therefore, we identify a significant knowledge gap in the application of TA in the public sector in general and, given the data generated during citizen digital engagement processes, particularly for the co-creation of public services as emphasized in several of the European's commission high-level policies (e.g., Berlin Declaration on Digital Society and Value-based Digital Government [14]) and frequently found on policymakers' agenda. Researchers increasingly recognize that public administration paradigms and theories will not suffice to understand the benefits and risks of co-creation [18].

In recent years there have been efforts to tighten the definition of co-creation and how it is similar to or differs from co-production [99]. Whilst co-production stems from scholarly work on the public sector (collaboration between public departments and citizens, e.g. Ostrom), co-creation is more recent and comes from the private sector, but is understood as the more encompassing term, referring to all kinds of citizen inputs in service [18].

As research on TA in this area is limited we hope to find my literature by using the term co-creation. Co-creation is the joint effort of citizens and public sector professionals in the initiation, planning, design and implementation of public services [19]. For this study we draw on the definition by [96], that is, co-creation is a process by which public and private stakeholders address a shared problem via a positive interchange of diverse types of information, resources, and skills. Different scholars identify several phases in the co-creation process. To address the research gap we identified, we adopt a model of the co-creation of public services based on three phases: *co-design, co-delivery,* and *co-evaluation* [57] and the operationalization of the phases as also outlined in [57]. The description of these three phases and the mechanisms they include allow us to study the available literature and to answer our research questions in the present study context (see Table 1). The *co-design* phase provides an essential frame for the conception and layout of the service that is to be designed and engages different stakeholders in the development of the specific public service. The key mechanisms of co-design are (1) consultation and ideation of service design elements enabling the citizens to share their opinions with the government, e.g., via social media, and (2) informing the citizens, i.e., equipping them with the data needed for informed decision-making [57]. *Co-delivery* (or co-execution) enhances the acceptance of the services through the involvement of citizens in the delivery of the public service as well as promotes communication between service providers and citizens and a more integrative user experience [92]. This phase is based on such key mechanisms as (1) crowdsourcing whereby government delegates an activity for (co-)execution to citizens to take advantage of their special abilities, talents, and expertise and (2) ecosystem embedding in which government integrates into the community via contributions such as academic alliances or community health workers. *Co-evaluation* phase, also known as co-assessment or co-monitoring [88], assesses the service after its delivery to learn from it or to adapt to it through possible prospective elements [69]. In this phase, citizen reporting or participatory open data are important mechanisms.

Table 1: Co-creation phases and mechanisms (adapted based on [57])

| Phase (P) | Mechanisms (M) |
|---|---|
| P1. Co-design | **M1_1. Consultation and ideation**: enable citizens to share their opinions and ideas with public administrations in an interactive dialogue, e.g., provide information and received advice using social media and tools such as chatbots |
| | **M1_2. Informing**: public administrations use data mining and mapping prior to providing citizens with the data they need in order to make informed decisions |
| P2. Co-delivery | **M2_1. Crowdsourcing**: public administrations delegate an activity for (co-)execution to citizens to take advantage of their special abilities, talents, and expertise |
| | **M2_2. Ecosystem embedding**: public administrations integrate themselves into the community via contributions, such as academic alliances, community health workers |
| P3. Co-evaluation | **M3_1. Citizen reporting**: enable citizens to provide their feedback to the public administrations, e.g., sentiment in social media. |
| | **M3_2. Participatory open data**: proactive information dissemination by public administrations and citizens |

We suggest that TA contributes to the following taxonomy of public values [91]: economic value (output of public administration), administrative value (procedural perspective), societal value (societal perspective), and citizen value (individual perspective).

# 3 RESEARCH APPROACH

## 3.1 Literature selection and collection

We use the systematic literature review (SLR) and content analysis method proposed by [20, 101] for the methodological analysis and synthesis of the relevant literature and for answering research questions. The purpose of the SLR is to understand the current knowledge base of how and where TA provides support in the co-creation phases of public services and what value it can provide for public services. In the literature selection and collection process, three steps were conducted to cover the research objective comprehensively. In the first stage, the selection of the databases and the subsequent systematic literature search, collection and preliminary processing were performed to build a final set of papers for further analysis and synthesis. Five electronic databases were selected for the literature search: Scopus, IEEE Explore, ACM Digital Library, Web of Science, and Web of Science. The following search terms in English were applied to paper titles, abstracts, and keywords: ("Natural Language Processing" OR "Text Analytics" OR "Text Mining" OR "Textual Data") AND ("Public Service"). Attempts to add AND ("co-creation") to the list of keywords also resulted in zero results, so it was excluded from further search. We focus on the titles, abstracts, and keywords in English. Period 2019-2022 was chosen due to the following reasons: (1) the concept of co-creation in public services is relatively new, (2) significant growth of scientific and practical interest in the concept of co-creation and the use of TA techniques to support public services over the past few years.

In total, we identified 979 results from the five searched databases: Scopus (112 papers), IEEE (9), ACM Digital Library (95), Web of Science (19), and Springer (741). In the second step, we removed duplicates and non-English studies. As a result, 354 articles were excluded. In the third step, we identified eligible publications by title/abstract screening using the concept-centric approach [101]. For example, a broad list of text analytics methods and applications in public services was used to identify if the paper mentions potential public services co-creation support directly or indirectly. This lead us to identify 57 full-text articles. We then adopted a snowballing approach to enhance the search results. Several early publications, like from 2011 and 2014, were included in this step. Finally, 75 papers were included in the quantitative synthesis.

## 3.2 Data extraction and synthesis

*First,* to extract data from 75 full-text selected studies, we used a spreadsheet to record the metadata for each of the selected studies. This metadata includes descriptive information, text analytics techniques and addressed functionality, stakeholders involved in TA application; and main challenges that have been tackled by the TA application. To ensure the data extraction process quality, one of the experts extracted information from all included studies and then the other two experts reviewed the extracted information. The three experts are academics involved in research and consulting projects on text analytics, co-creation, and e-government.

*Second,* the research framework that aims to systematically organise the knowledge gained in the context of existing and potential applications of TA to support public service co-creation, were developed. For this, three independent experts iteratively analyzed and coded the raw data obtained through the previous steps:(1) identified TA techniques and their functionalities have been grouped into categories; (2) public service areas addressed in the studies were categorized; (3) TA techniques categories have been assigned to the co-creation phases according to coding scheme developed based on definitions provided in Table 1. In addition, the potential public value [91] expected from the use of TA in public service co-creation was derived. The example of the research framework coding scheme is presented in supplemental material[1].

Consistency of the expert results was ensured by implementing the following research approach: (1) during the first step, each expert categorized/assigned independently; (2) the experts discussed the validity of the results obtained and identified discrepancies; (3) the experts refined their results independently; (4) a second expert discussion was organized to guarantee the maximum consistency in results; and (5) a consensus was reached and final results approved by all experts. The systematic synthesis allowed us to group the identified TA techniques and their functionalities into categories and assign these categories to the co-creation phases (Table 1).

## 4 CHARACTERISTICS OF THE STUDIES

This section aims to present the main results of this study. Section 4.1 provides the results, including publication by year and country, categories of text analytics techniques, data sources used in the public service context, TA functionalities, and public services. Section 4.2 presents the potential public value to be gained from TA techniques, whilst Section 4.3 maps the results onto the co-creation phases.

### 4.1 Text Analytics techniques, data sources, functionalities, and public services

The first step was to extract the year of publication and country to identify when and where the studies on TA application in public services had been conducted. Regarding the distribution of the publication years, the majority of the articles were published in 2022 (33.3%), followed by 2021 (24.0%) and 2020 (18.65%). This shows that the trend has steadily been growing in the last years and can be explained by research and practical interest in developing solutions to support public services using state-of-the-art technology achievements. It also shows a move away from the New Public Management Paradigm and a renewed interest in the co-creation approach originally developed by Ostrom [78]. The analyzed studies were carried out in 33 different countries, although this number is likely to be higher as in 24 papers (32%), the country could not be identified. Interestingly, most of the research is conducted in the EU, UK, and Ireland (34.67%), followed by India and Indonesia (9.33%) and the USA and China (8%).

The reviewed studies allowed us to identify 28 standalone TA techniques, which we grouped into ten main categories used in the public service context. The majority of papers employ several TA techniques. Seven studies do not focus on a particular TA technique but provide an overview of specific topics, such as chatbots [1] or AI [43]. Figure 1a provides an overview and distribution of TA technique categories TA1-TA10. The most representative categories are TA1. *Machine learning-based techniques* (17.69%), such as text summarization, classification, clustering, association rules, and LSA; TA2. *Chatbot* (16.92%); and TA3. *Sentiment analysis* (13.08%) followed by TA4. *Topic modeling* and TA5. *Semantic Web and Linked Data* (10.77% each). The TA10. *Multilingual solutions* are the least represented category (2.31%). The overall popularity of these techniques in applied research and practice of TA conditions this distribution. A notable observation is that aside from the high potential and possibilities offered by state-of-the-art TA, a significant proportion of the studies uses manual and rule-based approaches, i.e., TA6. *Content analysis* category (10.00%), including discourse analysis, thematic analysis, and rule-based matching, surpassing the TA7. *Artificial neural networks* (6.92%). This can be explained by the lack of trust and the traditional, historically-driven nature of public services. The TA techniques are visualized by the data sources (DS) to which they are applied (see Figure 1b). Based on the figure, typical TA trends of data source usage can be identified, i.e., DS1. *User-generated content* (37.50%) and DS2. *Social media* (26.39%). Noted sector-specific data sources are DS4 and DS5 (8.33% each) and DS8 (1.39%).

---

[1] Research framework coding scheme

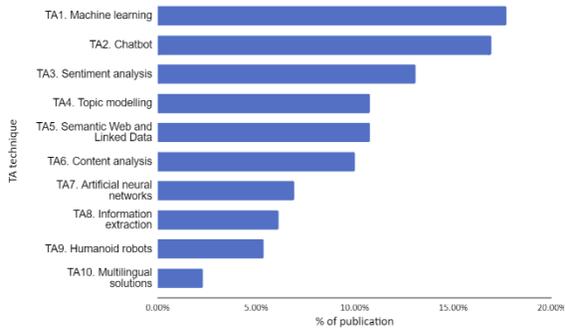
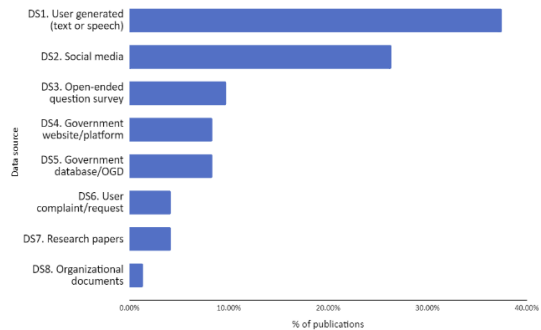

Figure 1a: Categories of TA techniques in public services    Figure 1b: Categories of TA data sources in public services

The identified TA techniques can provide specific functionalities in the context of public services. Thus, the functionalities described in the studies were also extracted and combined into ten categories F1-F10 (see Figure 2a), with the F1. *Automatic content analysis and extraction* (29.86%); F2. *Patterns identification* (24.31%); and F3. *Question answering* (18.06%) being the most prominent ones and F10. *Multilingual support* (1.39%) the least prominent one. The extracted functionalities reflect the specific demands of public services on the one side and the potential for the development of powerful cross-service solutions on the other. As can be seen from the public service distribution analysis, the most frequently recurring TA solutions were developed for a cross-service application (generic approaches, 37.33%). Figure 2b provides an overview of the identified 14 public services PS1-PS14, excluding the most frequent category P15. *Generic public service*. Most cases were identified in the health service area (14.67%), as the health and social public services require typical co-creation activities. For example, [104] developed a chatbot for supporting users with mental health conditions. [76] analyzed free-text feedback to detect the core determinants of patients' experiences in hospitals to provide recommendations for healthcare service improvement.

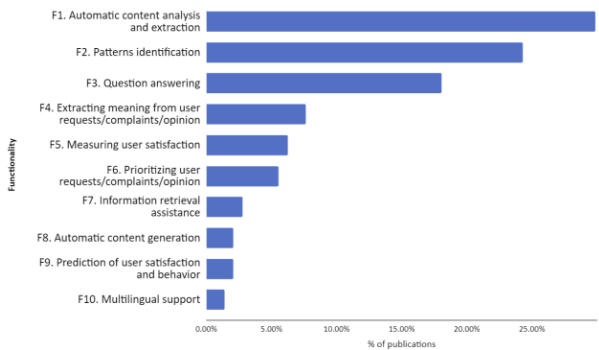
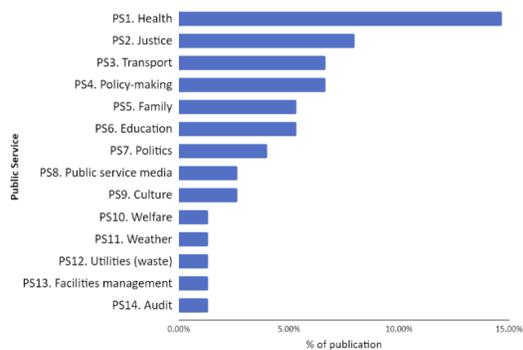

Figure 2a: Categories of TA functionalities in public services    Figure 2b: Public services

## 4.2 Co-creation phases supported by Text Analytics techniques

To answer the research question RQ1: *During which co-creation phases can TA support the development of public services?*, the extracted TA techniques have been assigned to the co-creation phases. The co-creation phases used in the study are based on the approach by [57]. Table 2 shows the number of articles mapped to the co-creation phases and mechanisms mentioned in Table 1. As our study results show (see Figure 4), the main focus of research is on the application of TA for public services to support the P1. *Co-design* phase (71.70%). Hereby, the P3. *Co-evaluation* phase is 23.58%, and only 4.72% of all research is aimed at supporting the P2. *Co-delivery* phase. It is important to note that some articles addressed two phases of co-creation and multiple mechanisms. Whereas the majority of papers (73%) address only one co-creation phase, a positive trend (27%) of involving TA techniques in two phases was noted. Overall, most phases supported by TA techniques separately are P1. *Co-design* (in almost 60% of papers) and P3. *Co-evaluation* (in 13.51% of papers). 20.27% of articles support the P1. *Co-design* and P3.*Co-evaluation* phases together. Several articles (6.76%) also address P1. *Co-design* and P2. *Co-delivery* phases together. Such a distribution appears appropriate due to the research character of the studies conducted in the papers, i.e., the suggested solutions are likely to be in the design and evaluation phase, as well as the analytic nature of TA. The most prominent co-creation mechanisms are M1_1. *Consulting & ideation* (28.38%), M1_2. *Informing* (20.27%), and M3_1. *Citizen reporting* (13.51%), which reflects the type of research and TA application specifics.

Regarding RQ2: *Which TA techniques best support digital co-creation?*, in the column "Text Analytics techniques/Source" of Table 2 we listed (in descending order) the TA techniques that were introduced in the analysed papers in the context of public services and have the theoretical potential for their application to support the public service *co-creation* in different phases. Accordingly, TA2. *Chatbot* shows a high potential to support a co-design phase, especially consulting & ideation. State-of-the-art chatbot technology allows a convenient, time-independent notification of citizens regarding relevant service updates and events and the development of constructive dialogue in the context of service improvement. Moreover, chatbots are powerful tools for searching and exploring various data sources, for example, open data [23], representing significant support for consulting & ideation. In this relation, TA9. *Humanoid robots* can be considered a more social, however not frequently used, physical alternative to chatbots. Addressing the limitations of intelligent interaction, [22] propose a humanoid robot with the self-learning capability for accepting and giving responses from people to be widely used in public services.

Further, TA1. *Machine learning*, including text classification, clustering, and summarization, is likely to support the co-design phase, particularly the mechanism "informing". These TA techniques are efficient in processing large amounts of text, extracting relevant information, and preparing it for further transmission, i.e., enabling informing citizens with a targeted message and not overwhelming them with large amounts of content. For instance, [47] used text mining of smart city initiatives texts to evaluate the status and progress of smart city development projects.

TA3. *Sentiment analysis* aims to extract the opinion and attitudes of citizens to specific services and/or related problems and issues. Thus, it is mainly used to support the co-evaluation phase, particularly citizen reporting. [67] designed the bureaucracy sentiment analysis application to analyze the opinions about bureaucratic services in Indonesia. Sentiment analysis is also used to support co-design informing, or to extract socially relevant information from large distributed data sources, such as the president's annual speech analysis, to derive sentiments and emergent topics [62]. In this sense, TA4. *Topic modeling* appears to be a second potentially used technique in the co-design phase for the mechanism "informing". [33] apply topic modeling to discover key topics from free text dispatcher observations in emergency medical dispatch. The analysis results can be provided to the interested authorities and citizens with the ultimate goal of positively impacting patient well-being and health services sustainability. Topic modeling also supports co-evaluation phase, citizen reporting, and user profile development in the context of citizen-government communication [34].

Another popular set of TA techniques TA5. *Semantic Web and Linked Data* can potentially find its application in assisting the co-design phase in both consulting & ideation and informing. These technologies provide a common structure and tools for integrating and reusing information [4], making it semantically interoperable and serving as a basis for many applications, for example, chatbots or expert systems. In this respect, several research directions emerged to enhance a so-called Core Public Service Vocabulary Application Profile (CPSV-AP), a standard European public service data model to facilitate public service catalog creation and semantic interoperability [7, 37]. Such vocabularies and ontologies are applied in developing various expert systems to facilitate informed decision-making (co-design, informing). For instance, [9] work on an open and automated legal expert system capable of providing any EU country's legal information based on the existing ontologies.

Due to trust and transparency limitations and ethical issues, TA7. *Artificial neural networks* have not yet found a wide application in the support of public service co-creation. Several possible cases were identified in support of the co-design phase, consulting & ideation. These cases are mainly found in challenging settings, such as expert systems developed for health insurance [49] or multilingual chatbots [6], and require sufficient training data. Artificial neural networks are also often combined with other TA techniques to develop a more robust and meaningful solution. For example, the research by [34] uses this TA to suggest user profiles based on the citizens' free-text requests. The observation that TA7 is used as a complementary TA in a mix of techniques is reflected in the articles that focused on two co-creation phases. For instance, developing a legal expert system to support legal officials in performing their duties by collecting and analyzing the information obtained from previous court cases demands the application of TA2, TA5, and TA7 [84, 94].

TA8. *Information extraction*, TA6. *Content analysis*, and TA10. *Multilingual solutions* are the less-used TA techniques that can be justified by the paper distribution in the sample and generic TA popularity. TA8 and TA10 are found in a co-design phase, consulting & ideation, implying a certain interactivity, whereas TA6, having a more analytical nature, is found in the co-design phase, informing. In the papers addressing two co-creation phases, these TA make part of a set of techniques potentially used in co-design and co-evaluation phases with various co-creation mechanisms: consulting & ideation, informing, citizen reporting, and open book government.

Table 2: Co-creation phases supported by Text Analytics techniques

| Co-creation Phases | Co-creation Mechanisms | Text Analytics Techniques / Source | % of papers |
|---|---|---|---|
| *One co-creation phase* | | | |
| P1. Co-design | M1_1. Consulting & ideation | TA2. Chatbot [7, 16, 80, 104, 23, 27, 30, 38, 44, 59, 64, 73]<br>TA9. Humanoid robots [6, 7, 22, 30, 64]<br>TA5. Semantic Web and Linked Data [36, 37, 80, 82, 89]<br>TA7. Artificial neural networks [6, 22, 30, 49]<br>TA10. Multilingual solutions [16, 30]<br>TA8. Information extraction [49]<br>TA1. Machine learning [100] | 28.38 |
| P1. Co-design | M1_2. Informing | TA1. Machine learning [2, 13, 47, 63, 83, 87, 90]<br>TA4. Topic modeling [13, 33, 53, 62, 65, 97]<br>TA3. Sentiment analysis [13, 53, 62, 83]<br>TA5. Semantic Web and Linked Data [53, 83]<br>TA2. Chatbot [43, 50]<br>TA6. Content analysis [41] | 20.27 |

| Co-creation Phases | Co-creation Mechanisms | Text Analytics Techniques / Source | % of papers |
| --- | --- | --- | --- |
| P3. Co-evaluation | M3_1. Citizen reporting | TA3. Sentiment analysis [29, 46, 55, 67, 86]<br>TA4. Topic modeling [34, 54, 55, 86]<br>TA1. Machine learning [28, 34, 54]<br>TA7. Artificial neural networks [34] | 13.51 |
| P1. Co-design | M1_1. Consulting & ideation;<br>M1_2. Informing | TA2. Chatbot [1, 25, 60, 72, 81, 93]<br>TA1. Machine learning [26, 42] | 10.81 |
| *Two co-creation phases* | | | |
| P1. Co-design;<br>P3. Co-evaluation | M1_2. Informing;<br>M3_1. Citizen reporting | TA4. Topic modeling [52, 74–76]<br>TA1. Machine learning [51, 75, 76, 103]<br>TA3. Sentiment analysis [51, 74]<br>TA7. Artificial neural networks [51]<br>TA5. Semantic Web and Linked Data [51] | 8.11 |
| P1. Co-design;<br>P3. Co-evaluation | M1_1. Consulting & ideation;<br>M1_2. Informing<br>M3_1. Citizen reporting | TA8. Information extraction [12, 85]<br>TA1. Machine learning [40]<br>TA2. Chatbot [98]<br>TA5. Semantic Web and Linked Data [12]<br>TA3. Sentiment analysis [12] | 5.41 |
| P1. Co-design;<br>P2. Co-delivery | M1_2. Informing;<br>M2_1. Crowdsourcing | TA2. Machine learning [9, 39, 84, 94]<br>TA5. Semantic Web and Linked Data [9, 39]<br>TA7. Artificial neural networks [94] | 5.41 |
| P1. Co-design;<br>P3. Co-evaluation | M1_2. Informing;<br>M3_2. Open book government | TA7. Artificial neural networks [70]<br>TA2. Machine learning [70, 102]<br>TA3. Sentiment analysis [102]<br>TA6. Content analysis [10]<br>TA8. Information extraction [10] | 4.05 |
| P1. Co-design;<br>P3. Co-evaluation | M1_1. Consulting & ideation;<br>M3_1. Citizen reporting | TA2. Chatbot [37]<br>TA3. Sentiment analysis [66]<br>TA1. Machine learning [37, 66] | 2.70 |
| P1. Co-design;<br>P2. Co-delivery | M1_1. Consulting & ideation;<br>M2_1. Crowdsourcing | TA2. Chatbot [31]<br>TA1. Machine learning [31] | 1.35 |

### 4.3. Potential public service values based on Text Analytics techniques

As mentioned above, most of the TA solutions have been introduced in the general public service context, allowing for applications that can be used across various services (cross-service). In this respect, the following question arises: RQ3: *Which public service values do TA techniques contribute to co-creation?* It is unclear how digitization (and TA as part of it) is creating public value. Public value may be defined in various ways, but relatively limited empirical studies offer recommendations on how public value is actually co-created [91]. In the digital government context, public value can be considered an outcome and by-product of investments in digitization; however, measuring or operationalizing the concept is not addressed well [79]. Similar to [91], in the study context, public value is defined as citizens' expectations towards public service. Building on the recent taxonomy of public values [91], in the study, we use the four types of public values: (1) economic value as efficiency gains enabled by TA (e.g., less government expenditure and less human resources, increased efficiency); (2) administrative value as a better delivery of public services (e.g., better rules and regulation, single line of communication, personalized public service); (3) societal value as benefits provided by public services to society and framed within the legal rules (e.g., increased trust in the public sector, happier customer, security); (4) citizen value as direct value for an individual using the service (e.g., transparency, reduction of administrative burden, increase in number of services delivered).

As the results of Table 3 show, the top three EV. *Economic values*, which is also the most represented group (52.94%) that can be potentially obtained by applying TA techniques are EV1. *Increased efficiency* (19.72%), EV2. *Time-saving for the authorities and businesses* (11.76%), and EV3. *Easier to contact with public service* (8.65%). The proactive identification of issues related to public service based on sentiment analysis [66] and text mining algorithms [40], implementing various chatbot services to immediately provide information to citizens 24/7 [93] increase the efficiency of the service and contribute to cost savings. When performing the analysis, multiple TA techniques could be mapped to a single value. For instance, chatbot services reduce the time public authorities need to resolve the most common and straightforward requests. TA techniques, including humanoid robots [22] and machine translation [30], contribute to a more pleasant and efficient contact provision of public service. The second most represented group is CV. *Citizen values* (5.88%), where CV1. *Transparency* (5.88%), CV2. *Reduction of administrative burden* (5.54%), and CV3. *Better services due to quality standards* (4.84%) are the top three. TA techniques are known for their capability of processing and extracting meaning from large amounts of text from multiple sources. Thus, automated identification of national implementations of European directives based on a large legal text corpus [70] can bring transparency on how far the directives are implemented at a national level. Various service automation solutions supported by chatbots [43] or assistance in filling the documents [7] will likely reduce administrative efforts and contribute to better quality service. In the third group, SV. *Societal values* (19.38%), the most represented value is SV1. *Satisfied public service user* (16.61%). In the long run, an overall goal and the outcome of any public should be user satisfaction. Hence, implicitly, the majority of TA techniques, like the aforementioned chatbots, sentiment analysis, and machine learning-based TA, should aim to achieve this goal. Finally, the least represented group is the AV. *Administrative values* (1.73%). This is a rather "hands-on", practical set of values, which makes it challenging to derive the potential of using TA techniques. For example, designing a chatbot service or investigating possible options of chatbot application for public service delivery [60] implies the revision, adaption, and improvement of existing administrative rules and regulations.

Table 3: Potential public service values based on TA techniques (%)

| Public service value | EV. Economic value | CV. Citizen value | SV. Societal value | AV. Administrative value |
| --- | --- | --- | --- | --- |
| EV1. Increased efficiency | 19.72 | | | |
| EV2. Time-saving for the authorities and businesses | 11.76 | | | |
| EV3. Easier to contact with public service | 8.65 | | | |
| EV4. Enhance the use of digitalization and new business models | 5.88 | | | |
| EV5. Less human resources | 5.54 | | | |
| EV6. Less government expenditure | 0.69 | | | |
| EV7. Cost-effective digitalization | 0.69 | | | |
| CV1. Transparency | | 5.88 | | |
| CV2. Reduction of administrative burden | | 5.54 | | |
| CV3. Better services due to quality standards | | 4.84 | | |
| CV4. Increase in number of services delivered | | 4.15 | | |
| CV5. Better access and understandability of services | | 3.46 | | |
| CV6. Personalized service delivery | | 2.08 | | |
| SV1. Satisfied public service user | | | 16.61 | |
| SV2. Increase trust in public service | | | 1.73 | |

| Public service value | EV. Economic value | CV. Citizen value | SV. Societal value | AV. Administrative value |
|---|---|---|---|---|
| SV3. Security | | | 0.35 | |
| SV4. Learning from each other | | | 0.35 | |
| SV5. Best practices & lessons learned | | | 0.35 | |
| AV1. Better rules and regulation | | | | 0.69 |
| AV2. Single line of communication | | | | 0.69 |
| AV3. Easier business compliance with rules and regulations | | | | 0.35 |

On the basis of all results of data synthesis, we propose the Research Framework for the use of TA techniques in public service co-creation (Figure 3). The process of co-creation can be split into several distinct phases, and in our study, we draw on the following three phases: co-design, co-delivery, and co-evaluation, and further informative aspects associated with each of these phases. The framework is suggested to bring together and specify the following aspects: (1) requirements and needs of the co-creation phases, (2) type of public services that need to be fulfilled by the use of TA techniques, (3) resources a public service can provide for developing TA-based solutions, (4) the functionalities TA can provide to support co-creation of public services, and (5) economic, administrative, societal and citizen values, the achievement of which, as we suggest, to a greater extent can be supported by the application of TA techniques.

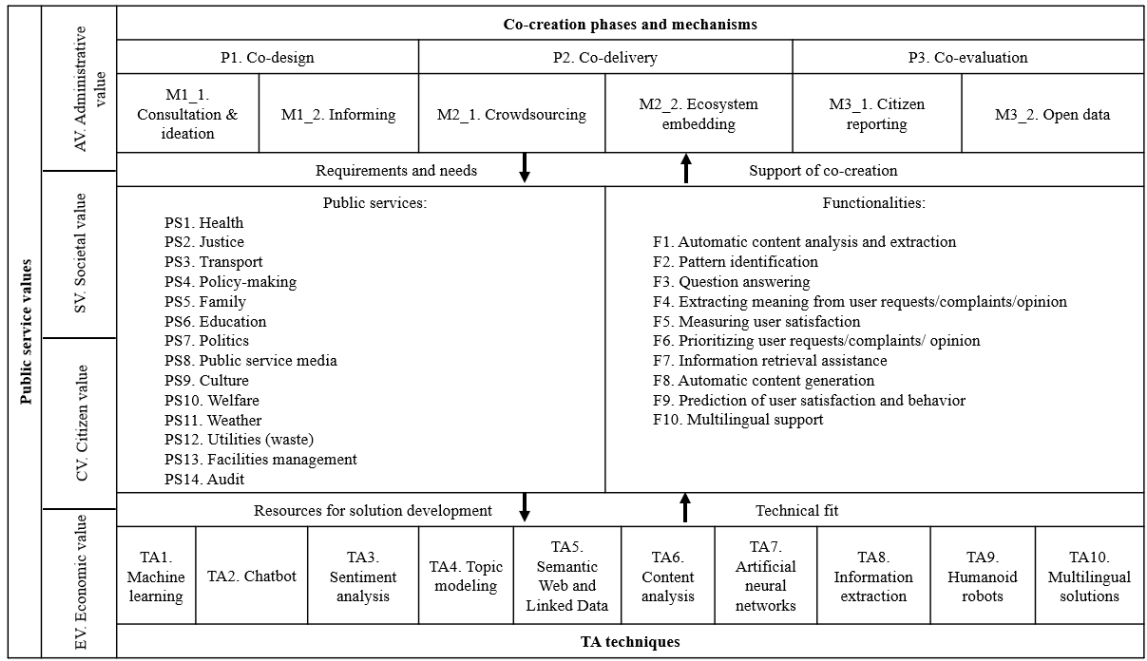

Figure 3. Research framework for Text Analytics techniques for the public service co-creation

## 5 DISCUSSION AND CONCLUSION

In the present work the framework for the use of TA techniques in public service co-creation is developed. The majority of the studies that build the basis of the proposed framework is academic or commercial research with a focus limited to the possibilities of application of TA to support public services. Accordingly, the mapping of TA techniques extracted from gathered literature to main co-creation phases and mechanisms was carried out by identifying the existence of a theoretical potential for their application to directly support the process of public service *co-creation.* Our study makes some important *contributions*. First, it contributes to the emerging body of literature by increasing awareness of the recent status of using TA techniques to support public service. Second, from a practical perspective, our research framework promotes the concept and stimulates the strengthening of the role of TA techniques in supporting of public services co-creation process in all phases. Third, from policy-makers' and e-government managers' standpoints, our findings and the proposed framework can be used as a guideline in developing a strategy for designing user-centred public services.

Our study has *limitations* that need to be articulated: (1) TA application challenges and adverse effects on public service co-creation should be carefully studied; (2) the main requirements for policies and ethical implications of TA in public services co-creation should be established; (3) only English-speaking studies were included. We hope it would be possible to select a specific range of countries and analyze the policy data of each country directly. *Future research* may focus on the following directions: (1) research framework can be tested in various application scenarios to extend framework by more concrete structural or procedural aspects; (2) since sustainability is an increasingly important issue in the public sector, TA functionalities have to consider whether the support they provide can be used only in the development of sustainable public services or if they can also contribute to the development of sustainable co-creation processes and practices in public administrations; (3) necessary expertise and competences development required for using TA techniques in co-creation should be considered.

## ACKNOWLEDGMENTS

This project has received funding from the European Union's Horizon 2020 research and innovation programme under Grant Agreement 962563 (Project inGOV: Inclusive Governance Models and ICT Tools for Integrated Public Service Co-Creation and Provision).

180–184.

[50] Jones, B. and Jones, R. 2019. Public Service Chatbots: Automating Conversation with BBC News. *Digital Journalism*. 7, 8 (Sep. 2019), 1032–1053.

[51] Kim, N.R. and Hong, S.G. 2020. Text mining for the evaluation of public services: the case of a public bike-sharing system. *Service Business*. 14, 3 (Sep. 2020), 315–331.

[52] Kowalski, R., Esteve, M. and Jankin Mikhaylov, S. 2020. Improving public services by mining citizen feedback: An application of natural language processing. *Public Administration*. 98, 4 (Dec. 2020), 1011–1026.

[53] Lachana, Z., Loutsaris, M.A., Alexopoulos, C. and Charalabidis, Y. 2021. Clustering legal artifacts using text mining. *ACM International Conference Proceeding Series*. (Oct. 2021), 65–70.

[54] Lee, H.J., Lee, M. and Lee, H. 2018. Understanding public healthcare service quality from social media. *Lecture Notes in Computer Science (including subseries Lecture Notes in Artificial Intelligence and Lecture Notes in Bioinformatics)*. 11020 LNCS, (2018), 40–47.

[55] Lee, H.J., Lee, M., Lee, H. and Cruz, R.A. 2021. Mining service quality feedback from social media: A computational analytics method. *Government Information Quarterly*. 38, 2 (Apr. 2021).

[56] Lember, V., Brandsen, T. and Tõnurist, P. 2019. The potential impacts of digital technologies on co-production and co-creation. *https://doi.org/10.1080/14719037.2019.1619807*. 21, 11 (Nov. 2019), 1665–1686.

[57] Linders, D. 2012. From e-government to we-government: Defining a typology for citizen coproduction in the age of social media. *Government Information Quarterly*. 29, 4 (2012), 446–454.

[58] Liu, B. 2012. Sentiment analysis and opinion mining. *Synthesis Lectures on Human Language Technologies*. 5, 1 (2012), 1–184.

[59] Magnini, B., Not, E., Stock, O. and Strapparava, C. 2000. Natural language processing for transparent communication between public administration and citizens. *Artificial Intelligence and Law 2000 8:1*. 8, 1 (Mar. 2000), 1–34.

[60] Makasi, T., Nili, A., Desouza, K.C. and Tate, M. 2022. A Typology of Chatbots in Public Service Delivery. *IEEE Software*. 39, 3 (2022), 58–66.

[61] Medelyan, O., Witten, I.H., Divoli, A. and Broekstra, J. 2013. Automatic construction of lexicons, taxonomies, ontologies, and other knowledge structures. *Wiley Interdisciplinary Reviews: Data Mining and Knowledge Discovery*. 3, 4 (2013), 257–279.

[62] Miranda, J.P.P. and Bringula, R.P. 2021. Exploring Philippine Presidents' speeches: A sentiment analysis and topic modeling approach. *Cogent Social Sciences*. 7, 1 (2021).

[63] Moreira Valle, L., Giacomazzi Dantas, S., Guerreiro E Silva, D., Silva Dias, U. and Monteiro Monasterio, L. 2022. RegBR: A novel Brazilian government framework to classify and analyze industry-specific regulations. *PloS one*. 17, 9 (2022), e0275282.

[64] Mrsic, L., Mesic, T. and Balkovic, M. 2020. Cognitive Services Applied as Student Support Service Chatbot for Educational Institution. *Advances in Intelligent Systems and Computing*. 1087, (2020), 417–424.

[65] Muguro, J., Njeri, W., Matsushita, K. and Sasaki, M. 2022. Road traffic conditions in Kenya: Exploring the policies and traffic cultures from unstructured user-generated data using NLP. *IATSS Research*. 46, 3 (Oct. 2022), 329–344.

[66] Muktafin, E.H. and Kusrini, P. 2021. Sentiments analysis of customer satisfaction in public services using K-nearest neighbors algorithm and natural language processing approach. *Telkomnika (Telecommunication Computing Electronics and Control)*. 19, 1 (Feb. 2021), 146–154.

[67] Muliawaty, L., Alamsyah, K., Salamah, U. and Maylawati, D.S. 2019. The concept of big data in bureaucratic service using sentiment analysis. *International Journal of Sociotechnology and Knowledge Development*. 11, 3 (Jul. 2019), 1–13.

[68] Munné, R. 2016. Big data in the public sector. *New Horizons for a Data-Driven Economy: A Roadmap for Usage and Exploitation of Big Data in Europe*. (Jan. 2016), 195–208.

[69] Nabatchi, T., Sancino, A. and Sicilia, M. 2017. Varieties of Participation in Public Services: The Who, When, and What of Coproduction. *Public Administration Review*. 77, 5 (Sep. 2017), 766–776.

[70] Nanda, R., Siragusa, G., Di Caro, L., Boella, G., Grossio, L., Gerbaudo, M. and Costamagna, F. 2019. Unsupervised and supervised text similarity systems for automated identification of national implementing measures of European directives. *Artificial Intelligence and Law*. 27, 2 (Jun.